\documentclass[useAMS,usenatbib]{mn2e}  
  
\newcommand{\ltsima} {$\; \buildrel < \over \sim \;$} 
\newcommand{\simlt}  {\lower.5ex\hbox{\ltsima}}            % < over MMM 
\newcommand{\gtsima} {$\; \buildrel > \over \sim \;$} 
\newcommand{\simgt}  {\lower.5ex\hbox{\gtsima}}            % > over MMM 

\def\kms{km~s$^{-1}$}

\title[Emission line variability of RS Oph]
{Emission line variability of RS Ophiuchi   
\thanks{based on data from ESO (program 073.D-0724) and the ING Archive} }  
\author[Zamanov, Bode, Tomov, Porter ]  
{R. K. Zamanov$^{1,2}$\thanks{e-mail: rkz@astro.bas.bg; 
mfb@astro.livjm.ac.uk; tomov@astro.bas.bg; 
}    
M. F. Bode$^{1}$, N. A. Tomov$^{3}$, J.M. Porter$^{1}$ \\ 
$^{1}$ Astrophysics Research Institute, Liverpool John Moores University, 
Twelve Quays House, Birkenhead, CH41 1LD, UK \\ 
$^{2}$ Institute of Astronomy, Bulgarian Academy of Sciences,   
       72 Tsarighradsko Shousse Blvd., 1784 Sofia, Bulgaria \\  
$^{3}$ Institute of Astronomy, National Astronomical Observatory Rozhen, 
       POBox 136, 4700 Smolyan, Bulgaria  \\    
}  
  
\begin{document}  
  
\date{Accepted 12 July 2005. Received 2005 June 6}  
  
\pagerange{\pageref{firstpage}--\pageref{lastpage}} \pubyear{2004}  
  
\maketitle  
  
\label{firstpage}  
  
\begin{abstract}
 We report that the H$\alpha$ emission line of RS~Oph was strongly 
 variable during our 2004  observations on a time scale of 1 month. 
 The line consisted of both a double peaked central narrow component (FWHM$\sim$220 \kms)
 and a strongly variable broad one (FWHM$>$2000~\kms).
 The base of the H$\alpha$ line was very broad with FWZI$\approx$4600 \kms\ on all spectra from 1986 to 2004.  
 The variability of the broad component extends 	
 from  $-$2000 to +2000 \kms. Most probably this  is due to either blobs  ejected 
 from the white dwarf (with a typical blob mass estimated 
 to be $\sim$10$^{-10}$ M$_\odot$) 
 or a variable accretion disk wind.
 We also detected variability of the HeII$\lambda$4686 line on a time scale shorter than 1 day.	  
 The possible origin is discussed. 
\end{abstract}  
  
\begin{keywords}  
stars:individual: RS~Oph -- binaries: symbiotic --  
                binaries:novae, cataclysmic variables                   
\end{keywords}  
  
\section{Introduction}  

RS Ophiuchi (HD 162214) is a recurrent nova which underwent its last major
outburst in 1985 (see Bode 1987 and references therein). Following 
Dobrzycka \& Kenyon (1994), RS~Oph has a binary 
period of 460 days with orbital inclination to the line of sight of $30^0 - 40^0$, 
and spectroscopic ephemeris $T_0= JD2444999.9\pm29.3+460\pm10E$. 
The mass donor is a  red  giant of spectral type  K4-M0
(M{\" u}rset  \& Schmid 1999).
The nature of the hot component is unclear. Most probably it is a massive 
white dwarf with M$_{WD}\approx$1.4~M$_\odot$, 
accreting at $\dot{M}\ge$10$^{-8}$M$_\odot$~yr$^{-1}$ (Hachisu \& Kato, 2000).
Alternatively, it has been proposed to be a B-type shell star with highly variable luminosity,
which occasionally displays blue-shifted absorption features 
(Dobrzycka et al. 1996). 

% It deserves noting that a very massive white dwarf 
% with  M$_{WD}\ge$1.2~M$_\odot$ is very difficult to be formed 
% via normal stellar evolution (???????). It has to be formed in a binary. 

Broad, variable, emission components of the hydrogen lines 
were detected by Iijima et al. (1994), 
Anupama \& Miko{\l}ajewska (1999), and Tomov (2003). 
Variability of the He lines on a time scale of hours
was detected by Sokoloski (2003). The origin of these changes is a mystery.
Here we report that similar variability is also evident in our 
2004 observations and discuss its possible origin.  

\section{Observations} 
%%%------------------------------------------------------------
\begin{table*} %[!hb]
%\footnotesize
\caption{H$\alpha$ and HeII$\lambda$4686 observations of RS~Oph. In the table are given
number, date of observation, Julian Day, orbital phase,
radial velocity of the blue peak, central dip and red peak, full width at 
zero intensity (FWZI) of H$\alpha$, 
equivalent width of H$\alpha$, 
equivalent width of HeII$\lambda$4686 ,
the ratio between intensities of the blue and red peak, 
the radial velocity of the additional emission (V$_{emm}$), 
and the origin of the spectrum. 
The typical errors are  
$\pm$1 \kms\ for V$_{blue}$, V$_{dip}$ and V$_{red}$,
$\pm$250 km~s$^{-1}$ for FWZI,  
$\pm$5\%   for W$_{H\alpha}$, 
$\pm$20\%  for W$_{4686}$, 
$\pm$3\% for the V/R ratio, 
and $\pm$150 km~s$^{-1}$ for V$_{emm}$. The last column indicates the 
origin of the observations. }
\begin{center}
\begin{tabular}{cclccccccrlcr}
\hline
N: &  Date         &  JD &  $\phi$   &V$_{blue}$&V$_{dip}$  &V$_{red}$& FWZI	     & W$_{H\alpha}$ & W$_{4686}$ &  V/R & V$_{emm}$	 & origin \\
   &               &2\,400\,000+&     & \kms\   &\kms\     &\kms\   & km s$^{-1}$ & [\AA]	     & [\AA]	  &	 & [km s$^{-1}$] &  \\
\hline
    &              & 	       &       &     &	       &	&	&     &       &     &			  &	\\  
 1  & 1986-07-12   & 46624.4250& 0.53 &        &	  &	   & 4750  &  106  & ---   & 0.51&  $+$560:	     &  ING  \\    
 2  & 1986-07-13   & 46625.4868& 0.53 &        &	  &	   & 4940  &  102  & ---   & --- &  $+$740:	     &  ING  \\    
 3  & 1997-08-06   & 50667.3916& 0.32 &        &	  &	   & 4390  &  ---  & ---   & --- &  $-$1270	     &  ING  \\    
 4  & 1997-08-06   & 50667.3986& 0.32 &        &	  &	   & 4660  &  ---  & ---   & --- &  $-$1130	     &  ING  \\    
 5  & 2004-04-11   & 53106.3849& 0.62 &  -87.4  &   -48.9  &   +3.3& 4940  &  102  &$<$0.1 & 0.74&  $-$1440, $+$2065   &  ESO  \\	 
 6  & 2004-04-11   & 53106.3922& 0.62 &  -86.4  &   -47.9  &   +2.9& 4980  &  104  &$<$0.1 & 0.76&  $-$1420, $+$2060   &  ESO  \\    
 7  & 2004-06-05   & 53161.2236& 0.74 &  -92.8  &   -55.8  &   +0.7& 4160  &  125  & 1.9   & 0.53&  $+$1110	       &  ESO  \\    
 8  & 2004-06-05   & 53161.2311& 0.74 &  -92.8  &   -55.9  &   +2.0& 4210  &  125  & 3.2   & 0.54&  $+$1120	       &  ESO  \\    
 9  & 2004-06-06   & 53162.2840& 0.74 &  -90.7  &   -54.4  &   +2.4& 4530  &  127  & 0.7   & 0.60&  $+$1150	       &  ESO  \\    
 10 & 2004-06-06   & 53162.2916& 0.74 &  -89.8  &   -54.9  &   +2.9& 4300  &  120  & 0.6   & 0.59&  $+$1100	       &  ESO  \\    
 11 & 2004-08-31   & 53248.0434& 0.93 &  -91.0  &   -46.8  &   +2.9& 4890  &  110  & 0.2   & 0.86&  $-$1180, $+$1750   &  ESO  \\	
 12 & 2004-08-31   & 53248.0510& 0.93 &  -91.2  &   -45.6  &   +4.7& 4850  &  102  & 0.3   & 0.87&  $-$1225, $+$1740   &  ESO  \\	
    &              & 	       &       & 	&	       &	&	&	&	&   &			  &	  \\
\hline
\end{tabular}
\end{center}
\end{table*}

%%%------------------------------------------------------------

%%%------------------------------------------------------------
 %  RSOph.46624.4250.fits	 d01.fits   6504  6624  4754   106
 %  RSOph.46625.4868.fits	 d02.fits   6507  6635  4936   102
 %  RSOph.50667.3916.fits	 d03.fits   6512  6632  4388	59 
 %  RSOph.50667.3986.fits	 d04.fits   6512  6631  4662	58   
 %  RS-Oph.53106.38492584.fits   d05.fits   6508  6627  4936   102
 %  RS-Oph.53106.39221474.fits   d06.fits   6508  6624  4982   104
 %  RS-Oph.53161.22362489.fits   d07.fits   6510  6608  4159   125
 %  RS-Oph.53161.23118961.fits   d08.fits   6519  6607  4205   125 
 %  RS-Oph.53162.28405299.fits   d09.fits   6509  6623  4525   127
 %  RS-Oph.53162.29162341.fits   d10.fits   6511  6615  4296   120
 %  RS-Oph.53248.04349737.fits   d11.fits   6512  6621  4891   110
%
%;			        Ib   Ir
%;d01a      556.915	        9.1  18.0 
%;d02a      739.763
%;d03a     -1271.57
%;d04a     -1134.43
%;d05a     -1454.42
%;d05a      2065.41            15.7  21.1
%;d06a     -1408.70 2051.71    16.1  21.3
%;d07a      1014.04            15.6  29.2
%;d08a      1105.46            15.3  28.1
%;d09a      1151.17            15.9  26.7
%;d10a      1105.46            15.6  26.4 
%;d11a     -1180.14 1745.43    17.2  20.0
%;d12a     -1225.86            16.5  18.9 
%;d12a      1745.43
%
% 

We secured 8 spectra in 2004 using the ESO, La Silla, 2.2m telescope and the 
FEROS spectrograph.
FEROS is a fibre--fed echelle spectrograph, providing a resolution of 
$\lambda/\Delta \lambda=$48000, 
wide wavelength coverage from about 4000~\AA\  to 8000~\AA\  in one exposure 
and a high detector efficiency
% {\bf[or ``and a high throughput'' ??]} 
 (Kaufer et al. 1999).  

Additionally, we retrieved 
4 spectra covering the region of the H$\alpha$ line  
from the archive of the Isaac Newton Group of telescopes 
(ING). A journal of observations and the 
measured quantities are given in Table 1. 
%%%%%%%%%%%%%%%%%%%%%%%%%%%%%%%%%%%%%%%%%%%%%%%%%%%%%%%%%%%%%%%%%%%%%%%%%%%%%%%%%
 \begin{figure*}
 \mbox{}  
 \vspace{11.0cm}  
  \includegraphics{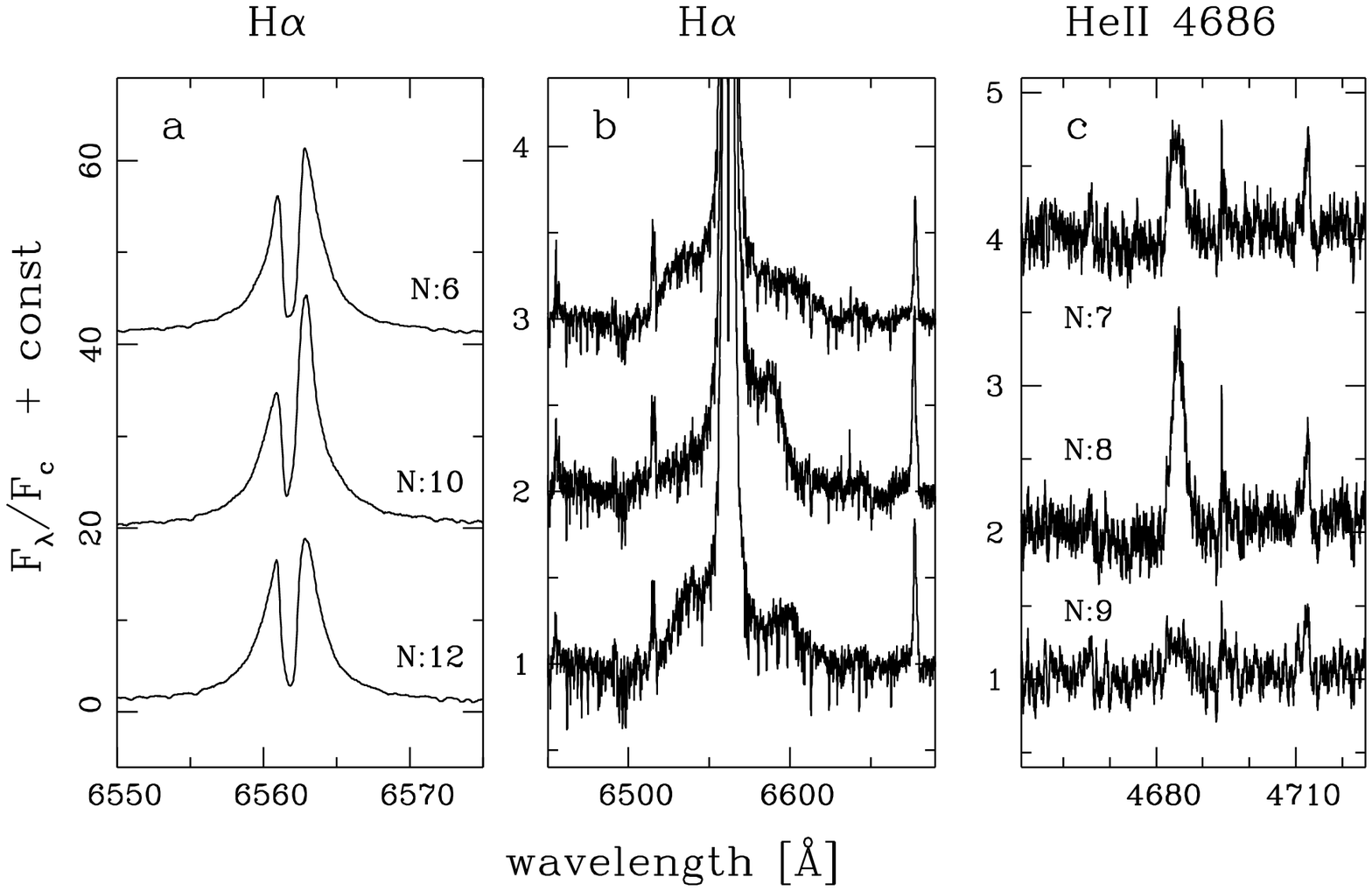}   
  \caption[]{The H$\alpha$ and the HeII$\lambda$4686 emission lines 
                        of RS~Oph (FEROS observations).
The number against each spectrum refers to the epoch of observation 
as given in Table 1. All spectra are normalized to the local continuum.
  
{\bf a)} The H$\alpha$ central double peaked component with  
FWHM $\sim$5\AA\ (220 \kms). 

{\bf b)} The same spectra as in a) but on  different scales, so  
the H$\alpha$ broad component is visible. 
 
{\bf c)} The HeII$\lambda$4686 line. The spectrum N:8 was obtained 10 minutes after N:7,
and N:9 - 24 hours later. 
These data demonstrate that the line is variable on a time scale $\le$1 day. 
  }		    
\label{RSOph-1}     
\end{figure*}	    
%%%%%%%%%%%%%%%%%%%%%%%%%%%%%%%%%%%%%%%%%%%%%%%%%%%%%%%%%%%%%%%%%%%%%%%%%%%%%%%%%%%%%%

\section{Spectral variability}

\subsection{H$\alpha$ emission line} 
The variability of the H$\alpha$ line 
is immediately apparent (see Fig.1 and Table 1). 
There are changes in the  blue and red peaks, 
as well as in the wings of the line. 
On all spectra the base of the H$\alpha$ line is very wide, 
the mean full width at zero intensity $\overline{\rm FWZI} =4650 \pm 300 $ km~s$^{-1}$
and the mean equivalent width of 
the H$\alpha$ emission line  $\overline{\rm W} = 112 \pm 11 $ \AA, 
which is about 1$\times$10$^{-11}$~erg~cm$^{-2}$~s$^{-1}$. 
%{\bf [Need to note that these are mean values?? You don't use the ``bar'' over the top anywhere else.]}
The ratio of the intensity of the blue and red peaks (V/R)
varies from 0.51 to  0.87. 
Our data suggest that the H$\alpha$  profile of RS~Oph is purely in emission
and consists of both a double peaked central narrow component 
(W$\sim$90 \AA, FWHM$\sim$200-250 \kms)
and a strongly variable broad one (W$\sim$20 \AA, FWHM$>$2000~\kms). 

When the spectra are obtained 
in the same night or in consecutive nights, the H$\alpha$ profiles are effectively identical. 
However,  changes are visible when the time difference is $\sim$1 month. 
To explore in more detail the velocity structure of the variability of the wings, 
assuming that the variability arises
purely from changes in emission, 
we subtracted at each epoch a minimum spectrum (artificially generated
by taking the minimum value from all normalized spectra at a given wavelength). 
The residuals are plotted in Fig.2b. 
As can be seen, the variability of the wings extends
from  $-$2000 to $+$2000~km~s$^{-1}$, with the most prominent additional emission 
around $-$1200 and  $+$1200~km~s$^{-1}$. 
On each spectrum we measured the radial velocity of the additional emission, 
which has equivalent width W$\le$7~\AA.
The data are given in Table 1 along with other parameters. 

A variable broad component was not detected in RS Oph's sister system 
T~CrB (see Stanishev et al. 2004; Zamanov et al. 2005). 
A similar component appeared however in the 1994 observations of CH~Cyg
(Tomov et al., 1996) and the velocity range of  $\pm$2000~km~s$^{-1}$,
was close to that of RS~Oph. However, 
the  time scale of appearance/disappearance was considerably shorter
($\sim$2 hours). 

%%%%%%%%%%%%%%%%%%%%%%%%%%%%%%%%%%%%%%%%%%%%%%%%%%%%%%%%%%%%%%%%%%%%%%
 \begin{figure}
 \mbox{}  
 \vspace{8.5cm}  
  \includegraphics{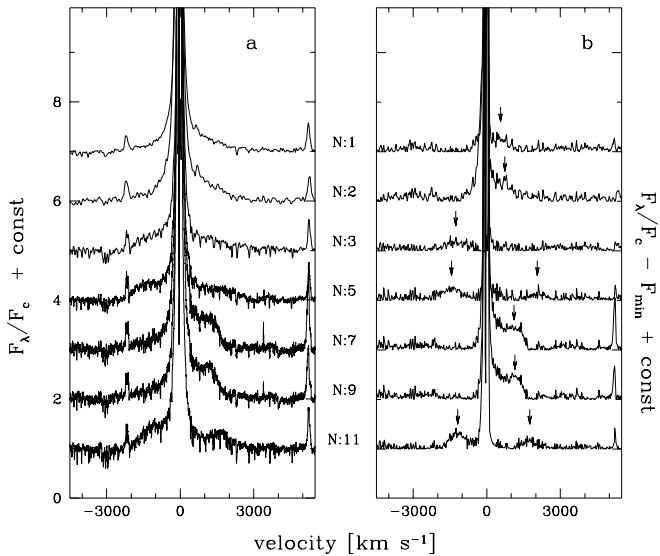}   
  \caption[]{{\bf a)} The RS Oph  H$\alpha$  profiles (normalized to the local
  continuum) for 7 different nights. 
The numbers again correspond to those in Table~1.
{\bf b)} After subtraction of the minimum spectrum, the additional emissions
are visible as bumps with W$\le$ 7 \AA. 
The arrows indicate their measured radial velocities as given in Table~1. 
  }		    
\label{RSOph-2}     
\end{figure}	    
%%%----------------------------------------------------------------------------------

\subsection{Time scale of HeII$\lambda$4686 variability}
Our spectra with numbers 7, 8 and 9  
show that the HeII$\lambda$4686 line is also variable
(Fig.1c and  Table 1). 
The spectrum N:8 was obtained 10 minutes after N:7, and N:9 - 24 hours later. 
This suggests that the intensity of HeII$\lambda$4686 changes on a time scale 
of minutes-hours (note  that we did not
detect H$\alpha$ variability on such a short time scale). 

The HeII$\lambda$4686 line indicates high excitation conditions in the gas. 
In the case of RS~Oph (see the next section) this can be related to a region
located close to the boundary layer between the accretion disk 
and the white dwarf (Sokoloski 2003). The changes of the intensity of this line
indicate changes in the region of its appearance which, in turn, 
is determined most probably by variable luminosity 
or/and temperature of the boundary layer.

%The variability of HeII$\lambda$4686 may be 
%connected with the boundary layer between the accretion disk and the
%white dwarf (Sokoloski 2003). 

%%%%---------------------------------------------------------------------------------
%%
%%   \clearpage
%%   \newpage
%%    \vskip 0.1cm
%%

\section{Discussion}
Our data suggest that the H$\alpha$  profile of RS~Oph is purely in emission
and consists of both a double peaked central narrow component and a strongly variable 
broad one.  The Balmer emission lines, appearing in the extended
envelopes of symbiotic stars usually have an ordinary nebular profile
with  typical FWHM$\sim$100-150 \kms,  for example AG Dra (Tomova \& Tomov 1999). 
FWHM increases to 
200 \kms\ only  during  the active phases. The basic mechanism determining their width
is turbulence in the gas. The central narrow emission of RS~Oph is very similar 
to the double peaked H$\alpha$ line of T~CrB and the two stars have practically 
the same FWHM of about 200-250 \kms. The H$\alpha$ line of T~CrB is supposed to be formed
in the outer part of an accretion disk around the hot compact
object (Stanishev et al. 2004). By analogy with this star we could assume that the 
narrow H$\alpha$ component 
of RS~Oph is formed also around  its compact object. This assumption, however meets some
difficulty with the fact that the peak of this component exceeds the level 
of the continuum by a factor of 20-25 like purely nebular lines of the symbiotic stars. 
This peak height differs from that one of T~CrB, which exceeds the level 
of the continuum by a factor of 2-6 only.

While the appearance and the variability of  the narrow component is relatively common 
in symbiotic stars (see also Ikeda \& Tamura 2004  and the references therein), 
appearance and variability 
of  broad component is detected in about $10$ objects up to now.
Fast ($\sim$1000 \kms) bipolar winds/jets were detected in Hen~3-1341 and StH$\alpha$~190 
(Tomov  2003). 
The jets in both systems appeared as satellite emission components on both sides of the strong 
HI Balmer and HeI emission lines, and they looked similar 
to the additional emission bumps in RS~Oph (Fig.~2b).
Similar profiles have been observed in the symbiotic nova RX~Pup 
(Mikolajewska et al. 2002) and AR~Pav 
(Quiroga et al. 2002). 

At quiescence, the hot components of RS~Oph and RX~Pup show activity (high states) 
characterized by the appearance of B/A/F shell-type features and variable 'false atmosphere' 
together with the broad and complex emission lines and are  related 
to the accretion flow/accretion-driven phenomena in these binaries.
Other systems that show similar behaviour are CH Cyg and MWC 560 (Mikolajewska et al. 2002, 2003).

Here, we will consider three possible origins for the broad component in RS~Oph: 

{\bf (1)} The first possible origin  is related to ejection of blobs of matter.
These blobs could be expelled by a rotating 
white dwarf magnetosphere (Tomov 2003 and references therein) or by a jet mechanism. 
The de-projected velocity of the ejection is $\sim$1500-4000 \kms,
having in mind V$_{emm}$ from Table 1 and assuming  
it is realized normal to the orbital plane, 
and also the orbital inclination is $30^0 - 40^0$.
This velocity is practically the same as ejection velocities of up to 3800 \kms seen during 
the 1985 nova outburst (Bode 1987 and references therein). 

{\bf (2)}
Variable disk winds with similar (to RS~Oph and CH~Cyg) velocities 
were detected  in a few cataclysmic variables 
and  the whole wind can even turn on and off. 
The terminal velocity of the wind in the  BZ~Cam system is v$_t \sim$3000~km~s$^{-1}$ 
and the time scale of the variability is 30-40 min (Ringwald \& Naylor 1998). Indeed, their 
profiles bear a noticeable similarity
to those in RS Oph at some epochs.
For  Q~Cyg  the terminal velocity is v$_t \sim$1500~km~s$^{-1}$  and 
the events last about 1.5 hours (Kafka et al. 2003). 
These time scales
are similar to  those of the Balmer line variability of CH~Cyg but not 
to RS~Oph. It is possible that the disk wind of RS~Oph varies on longer time
scales.

{\bf (3)} 
If the variable broad component of RS~Oph originates in an asymmetric disk, 
then a Keplerian velocity of 1000~km~s$^{-1}$  requires a distance from
a 1.4~M$_\odot$ WD of about 2$\times$10$^{10}$ cm. 
At that distance the Keplerian period is 20 min, which is considerably 
different from the observed time scale of H$\alpha$ variability of RS~Oph. 

All of the mechanisms considered are related to the loss of mass by the system.
Whenever such mass-loss occurs, a question  about its quantitative
estimate arises. A quantitative estimate, however, is possible in those
cases where the phenomenon can be related to some geometrical
model. However, we have no geometrical model of loss of mass which can be related
to the irregular form of the profile of the broad component. The velocity
distribution of this line proposes movement of discrete regions (blobs
of matter) in the emitting environment (Fig.2). That is why we tried to 
obtain a rough estimate of
the mass of one ``average" blob on the basis of its emission and supposing that
it is optically thin and has also a constant density within some limits. 
From Fig.1b and Fig.2, the luminous flux in H$\alpha$ 
of such a blob  is about 3$\times$10$^{-13}$ erg~cm$^{-2}$~s$^{-1}$ 
(W$\approx$ 3\AA). This flux was corrected for interstellar reddening using 
E(B$-$V)=0.73 (Snijders 1987; Hachisu \& Kato 2000, 2001) and 
the extinction law by Seaton (1979). A corrected flux of 
1.602$\times$10$^{-12}$ erg~cm$^{-2}$~s$^{-1}$ was obtained.

For calculation of the emission measure of this blob, data about the distance 
to the RS~Oph system, the H$\alpha$ recombination coefficient and the helium
abundance are needed. For the distance to the system we adopted 0.6 kpc according to
Hachisu \& Kato (2000, 2001). 
%{\bf[Note that Bode 1987, pp241-242 gives $d=1.6\pm0.3$ kpc from 
%a variety of measures. Is it best to use this value and scale the resulting blob mass accordingly??]}. 
It is not possible to obtain the electron temperature and the
electron density in the region where the H$\alpha$ broad component is emitted
from observation since we have no indication about the appearance in this 
region of certain lines giving information for those parameters. We
will suppose that the electron temperature is 20000 K  and the electron 
density is comparatively high being in the range 10$^8$-- 10$^{10}$ cm$^{-3}$. Then we used recombination
coefficients of 5.956$\times$10$^{-14}$ cm$^3$~s$^{-1}$ and 6.336$\times$10$^{-14}$ cm$^3$~s$^{-1}$ for case B, corresponding to these temperature
and densities (Storey \& Hummer 1995). We also adopted a helium abundance of 
0.1 which is thought to be typical of symbiotic nebulae (Vogel 1993;
Vogel \& Nussbaumer 1994).

For calculation of the emission measure we need to know the state of 
ionization of helium in the emitting region. The HeII$\lambda$4686 line
is absent in the spectrum (Dobrzycka et al. 1996) or is weak (Fig.1) 
during the quiescent state of the system. This means that singly 
ionized helium is dominant in the circumbinary nebula during this 
state. According to the suppositions presented however, the broad 
H$\alpha$ component is emitted in a region placed in the close vicinity
of the hot object in this system where probably helium is mostly doubly 
ionized. That is why here we will assume the state
of ionization to be He$^{++}$ in the emitting region.

Using a flux of 1.602$\times$10$^{-12}$ erg~cm$^{-2}$~s$^{-1}$ 
and density of 10$^8$-- 10$^{10}$ cm$^{-3}$, we obtain the emission measure
of one ``average" blob of matter (with a constant density) of 
3.19$\times$10$^{56}$-- 3.00$\times$10$^{56}$ cm$^{-3}$. 
We can also calculate the mass of the blob, adopting the
parameter $\mu$ of 1.4 (Nussbaumer \& Vogel 1987), determining the mean molecular
weight $\mu m_{\rm H}$ in the nebula. The mass is therefore obtained as
3.8$\times$10$^{-9}$-- 3.5$\times$10$^{-11}$ M$_\odot$. 
[Note that Bode (1987), gives $d=1.6\pm0.3$ kpc from 
a variety of measures. If we adopt this longer distance,
we  obtain mass of the blob 
2.7$\times$10$^{-8}$-- 2.5$\times$10$^{-10}$ M$_\odot$]. 

\section{Conclusions}

We report that the  H$\alpha$ emission of RS~Oph was strongly variable during
our 2004 observations on time scales of $\sim$1~month. No variability was
detected on time scales of $\le$1~day.   
The line was always very wide at its base (FWZI$\approx$4600~km~s$^{-1}$)
on all spectra from 1986 to 2004, consisting of narrow and broad emission 
components. Variable emission is detected at velocities up to
$\pm$2000~km~s$^{-1}$. 
Most probably this variable emission  is due to either blobs ejected 
from the white dwarf or a variable accretion disk wind.
The approximate mass of one "average" ejected blob of matter, suggested by the variability
of the  broad emission component is   
3.8$\times$10$^{-9}$-- 3.5$\times$10$^{-11}$ M$_\odot$. We also detected variability of 
HeII$\lambda$4686 line on time scales $<$1 day. 
To understand more fully the nature of the emission line variability  of RS~Oph
we need to acquire a set of spectra  with time resolution from minutes to days.

\section*{Acknowledgments}  
This note has made use of  IRAF, MIDAS and  Starlink.  
RZ is supported by a PPARC Research Assistantship  
and MFB is a PPARC Senior Fellow. 
We would like to thank the referee for the useful comments.
We also acknowledge the vital contribution made by Dr John Porter 
to this work, which was underway at the time of his tragic death.

\end{document}